\documentclass{article}
\usepackage{url}

\usepackage{graphicx}
\usepackage[pdftex,colorlinks=true,urlcolor=blue,citecolor=blue,anchorcolor=black,linkcolor=black,bookmarks=false]{hyperref}
\usepackage{natbib}

\title{Accounting for environmental awareness in wheat production through Life Cycle Assessment}
\author{Gianfranco Giulioni\footnote{University ``G. D'Annunzio'' of Chieti-Pescara, Italy, Department of Socio-Economic, Management and Statistical Studies, email: gianfranco.giulioni@unich.it (corresponding)} \and Edmondo Di Giuseppe\footnote{National Research Council (CNR), Italy, Institute for BioEconomy, \\ \hspace*{5mm} email: edmondo.digiuseppe@cnr.it, arianna.dipaola@cnr.it} \and Arianna Di Paola$^\dag$}
\begin{document}
\maketitle              % typeset the title of the contribution

\begin{abstract}
This paper presents a modeling framework for simulating the decision-making processes of artificial farms populating an agent-based model for the Italian wheat production system. The decision process is based on a mathematical programming model with which farms (i.e., agents) decide the target yield (production per hectare) and the mix of inputs needed to obtain such production, namely 1) fertilizers, 2) herbicides, and 3) insecticides. 
The environmental impacts of conventional production practices are assessed through a Life Cycle Assessment (LCA), using the ReCiPe 2016 methodology at the Endpoint level. Agents are made aware of the environmental consequences of their choices through two indicators: Disability-Adjusted Life Years (DALYs), which capture human health impacts, and the number of species lost per year, reflecting impacts on ecosystems. By internalizing this information, agents can make more balanced and sustainable production decisions.

\vskip3mm
\noindent\textbf{Keywods:} farm crop management, mathematical optimization, yield gap, ReCiPe methodology, agent-based model
\end{abstract}

\section{Introduction}
\label{sec:intro}

Wheat plays a central role in the global food system, serving as a staple crop for over one-third of the world’s population. Its adaptability to diverse climates, high yield potential, and nutritional value make it a key commodity for ensuring food security, particularly in Europe, Asia, and North Africa (\cite{FAO2021}). Beyond its role as a dietary staple, wheat production is a significant economic activity that supports rural livelihoods and contributes substantially to global agricultural trade.

Despite its critical role in global nutrition and the economy, wheat cultivation presents several environmental challenges. The intensive use of chemical fertilizers---particularly nitrogen-based compounds---contributes to nitrous oxide (N$_2$O) emissions, a greenhouse gas with a global warming potential nearly 300 times greater than carbon dioxide (\cite{IPCC2021,Snyder2009}). Moreover, wheat monoculture practices can lead to soil degradation, biodiversity loss, and increased vulnerability to pests and diseases, often resulting in a greater dependence on synthetic agrochemicals (\cite{Tilman2002}).

An emerging body of literature highlights the importance of enhancing environmental awareness among farmers as a key factor in reducing these negative impacts. Environmentally conscious agricultural practices---including conservation tillage, crop diversification, integrated pest management, and organic farming---are increasingly adopted by farmers who recognize the long-term benefits of ecological stewardship (\cite{Pretty2018}). Additionally, participation in agri-environmental schemes and sustainable certification programs has grown, supported by policy incentives and knowledge transfer initiatives (\cite{EU2020}).

Understanding the environmental impacts of wheat production and promoting sustainable farming practices are urgent priorities in the context of climate change and the depletion of natural resources. This paper aims to explore the environmental footprint of wheat cultivation and assess how environmental awareness among producers influences the adoption of sustainable practices. By integrating agronomic, ecological, and socio-economic perspectives, the study contributes to ongoing efforts to align wheat production with global sustainability goals.

In this work, we build an analytical crop management model to identify the quantities of three key inputs for wheat production: plant nutrients, herbicides, and insecticides (Section \ref{sec:inputs}). The model is calibrated using Italian agricultural data (Section \ref{sec:data}). Finally, we assess the impact in terms of human health and ecosystem quality using the Life Cycle Assessment (LCA) ReCiPe 2016 methodology (Section \ref{sec:lca} and \ref{sec:results}).
The ultimate goal of the ongoing research is to develop an agent-based model where agents (farmers) are informed about the environmental impact of their decisions. They, therefore, will be endowed with a feedback loop that can gradually bring them to revise their action in a more sustainable way. These aspects are discussed in the final Section.

%---------
\section{Modeling farm inputs decision}\label{sec:inputs}

\subsection{Farm crop management}
Modeling a farm's input decisions belongs to the broader field of farm management. Farm management has several aspects:
financial management, crop and livestock management, equipment management, labor management, and risk management (see \cite{kayetal2020} or \cite{farmmanweb}).
Since we specialize in wheat production, we will build on the tools used in the crop management field.
In particular, we are interested in modeling a situation in which a product (wheat) is produced using several inputs.
This choice is typically analyzed by applying economic principles (\cite{kayetal2020}, Chapter 8, p. 144). Indeed, the problem of choosing an input combination is a mathematical minimization, i.e., the farmer selects the input combination that minimizes costs. The dual problem of cost minimization is profit maximization (see \cite{carpentier_atal_2015} for a review of economic modeling of agriculture production). We analyze the problem of maximizing profit for one hectare of wheat, which is generally posed as follows:
\begin{equation}
\pi=p_wy(x_1,x_2,...)-\sum_i p_{x_i}x_i
	\label{eq:profit}
\end{equation}
where $y$ is yield per hectare, $x_i$ are inputs per hectare, $p_w$ is the price of wheat and $p_{x_i}$ are the inputs prices.
\\
In this work, we consider fertilizers, herbicides, and insecticides as inputs for wheat production.
A key role in the economic modeling of input combination choice is input substitution
(\cite{kayetal2020} chapter 8). The input substitution degree has been studied for several decades (see, for example, chapter 5 in \cite{headyetal1963}).
When inputs can be considered substitutes, the Cobb-Douglas or the CES (constant elasticity of substitution) are used as functional forms for $y(x_i)$.
Substitution commonly occurs between acreage and other production inputs; that is, it is possible to obtain the same production, for example, by increasing acreage and reducing fertilization.
\\
In this work, however, we propose a novel modeling strategy based on two key assumptions. First, since the analysis is conducted at the per-hectare level, we assume a zero degree of substitutability among inputs. This is justified by the agronomic observation that each input plays a specific and non-overlapping role in supporting plant health. Consequently, we model the production process using a Leontief-type production function, where yield depends on the most limiting input.
Second, we incorporate the concept of the yield gap—defined as the difference between potential and actual yield—as a guiding principle of farmer behavior. We assume that farmers aim to close this yield gap as much as possible, subject to profit maximization constraints.
To our knowledge, the use of the yield gap concepts is a novelty in farm microeconomics modeling. We therefore introduce this concept in the following section. 

\subsection{Yield-gap}

This approach starts by identifying the potential yield, i.e., the maximum yield obtainable depending on solar radiation, temperature, atmospheric CO2, and genetic traits. These features govern the length of the growing period. Thus, the potential yield is location-specific for several factors, especially the climate (see \cite{FISCHER20159} and \cite{yielgapclimatalk} for more detailed definitions and explanations).
The farm's realized yield is lower than its potential yield, and the difference between the potential and the actual yield is the yield gap. 
The yield gap is caused by limiting factors, such as water and nutrient availability, and reducing factors, including weeds, pests, diseases, and pollutants. Usually, a farm's yield does not exceed 80\% of the potential yield. Therefore, the concept of exploitable yield gap is introduced as the difference between the 80\% of the potential yield and the realized yield (see \cite{VANITTERSUM20134} page 5-6).
Several agronomic interventions—such as optimized fertilization, improved pest control, or precision agriculture—can help narrow this gap. For an overview of management practices with the most significant impact on yield, refer to Table 2 in \cite{DEVKOTA2024127195}.

It is important to note, however, that the yield gap is typically assessed ex post, at harvest time. In contrast, farm input decisions must be made ex ante, before the growing season. To address this asymmetry, the model developed in the next section introduces the concept of target yield—the yield level that the farmer aims to achieve, which balances the yield increase with input costs to maximize profit. 

\subsection{A model}\label{subsec:model}

We established a model based on the yield gap concept as a tool for informed management decisions.
The model identifies stress factors and suggests actions to alleviate them. Each production input can relieve one specific stress factor. Therefore, the farmer's action consists in weighing out the input quantity.
Let us index stress factors by $i$. 
We denote the conditional yield with $ y_i$, which is the yield obtained when only the stress factor $i$ is binding. 
Our first step is to set up a functional form for $y_i$.
Let us denote the potential yield with $\bar{y}$. In addition, we define $s_i\in(0,1)$ to identify the share of the potential yield lost due to the stress and $x_i$ as the strength of the measure taken to counteract the stress. % intensity of the stress
We also define $g_i(x_i)\in(0,1)$ as a function of $x_i$ that gives the effectiveness of the undertaken measure. Normally, $g_i$ is increasing in $x_i$ until it reaches a ceiling. We adopt the functional form $g_i(x_i)=1-e^{-\lambda_i x_i}$ having the just mentioned properties.
We further introduce the maximum share of yield that can be recovered at the maximum effectiveness of the measure. Let us identify it with $\bar{s}_i$.
\\
Under these definitions, the conditional yield can be written as
\begin{equation}
y_i(x_i)=\bar{y}[(1-s_i)+\bar{s}_i(1-e^{-\lambda_i x_i})]
	\label{eq:condy}
\end{equation}

When several stress factors are binding, the realized yield corresponds to that of the most binding stress factor: $y=\min(y_i)$.
As mentioned above, in the economic theory of production this is referred to as the Leontief-type function. Its main feature is that relieving one stress factor can be ineffective due to the constraints imposed by the other stress factors.
The optimal strategy in this case is to level out the conditional yields: $y_i=\hat{y}$.
\\
Using equation (\ref{eq:condy}), the $y_i=\hat{y}$ condition can be written as
\begin{equation}
	\bar{y}[(1-s_i)+\bar{s}_i(1-e^{-\lambda_i x_i})]=\hat{y}
	\label{eq:condyeqhaty}
\end{equation}
Solving $x_i$ we get:
\begin{equation}
	\hat{x}_i=-\frac{1}{\lambda_i}\ln\left(\frac{(1+\bar{s}_i-s_i)\bar{y}-\hat{y}}{\bar{s}_i\bar{y}}\right)
	\label{eq:hatx_i}
\end{equation}

With this result, we can go to the profit function (equation \ref{eq:profit}), which in our case is
\begin{equation}
	\pi=p_w \min(\hat{y}_i)-\sum_i p_{x_i}\hat{x}_i
	\label{eq:profitl}
\end{equation}
Because all the $\hat{x}_i$ deliver a yield equal to $\hat{y}$, we have $\min(\hat{y}_i)=\hat{y}$ and equation (\ref{eq:profitl}) simplifies to:
\begin{equation}
	\pi=p_w \hat{y}-\sum_i p_{x_i}\hat{x}_i
	\label{eq:profitlminfree}
\end{equation}
Remembering that $\hat{x}_i$ depends on $\hat{y}$, the whole profit function depends on $\hat{y}$. Therefore, the farmer's problem is to maximize profit with respect to $\hat{y}$:  
\[
	\max_{\hat{y}}\pi=p_w\hat{y}-\sum_ip_{x_i}\left[-\frac{1}{\lambda_i}\ln\left(\frac{(1+\bar{s}_i-s_i)\bar{y}-\hat{y}}{\bar{s}_i\bar{y}}\right)\right]
\]
The first order condition (FOC) for a maximum is:
\[
	p_w-\sum_ip_{x_i}
	\frac{1}{\lambda_i(1+\bar{s}_i-s_i)\bar{y}-\lambda_i\hat{y}}=0
\]
The FOC can be solved by numerical methods. Let us denote the solution with $\hat{y}^*$. Plugging $\hat{y}^*$ in equation (\ref{eq:hatx_i}), we obtain the optimal level of each input $\hat{x}^*_i$.

The one-stress-factor case can help with understanding because of its analytical solution. In the one stress factor case, we can drop the $i$ subscript from equations, and the sum symbol is not needed. Solving the farmer's maximization problem in this case, we obtain: 
$\hat{y}^*=(1+\bar{s}-s)\bar{y}-p_x/(p_w\lambda)$
and plugging into equation (\ref{eq:hatx_i}) we get the optimal input level:
$\hat{x}^*=-\frac{1}{\lambda}\ln\left(p_x/(p_w\lambda\bar{s}\bar{y})\right)$.

\section{Application to Italy} \label{sec:data}

The model in the previous section uses several parameter ($\bar{y}$, $\lambda_i$, $s_i$, $\bar{s}_i$), that we estimate using real world data. In this section, we provide a description of the database used for this purpose and details on identifying parameter values.  

\subsection{Database}\label{subsec:data}
The data on which we base our analysis are collected by the Council for Agricultural Research and Agricultural Economy Analysis (CREA) and recorded in the RICA dataset. The RICA acronym comes from the French expression “Réseau d’Information Comptable Agricole”, better known as “Farm Accountancy Data Network” (FADN). 
After inspecting the Italian RICA contents, the following variables were selected for the wheat cultivation of each farm:
\begin{itemize}
	\item cultivated hectares (ha);
	\item hours of tractor use per ha;
	\item fertilizers (kilograms of Nitrogen, Phosphorus, and Potassium per ha);
	\item pesticides (toxicity level and quantity of herbicides, insecticides, fungicides).
\end{itemize}

On the basis of the listed variable, we consider plant nutrition, weeds, and insects as stress factors.

\subsection{Calibration and estimation}
\begin{figure}[!ht]
	\includegraphics[scale=0.4]{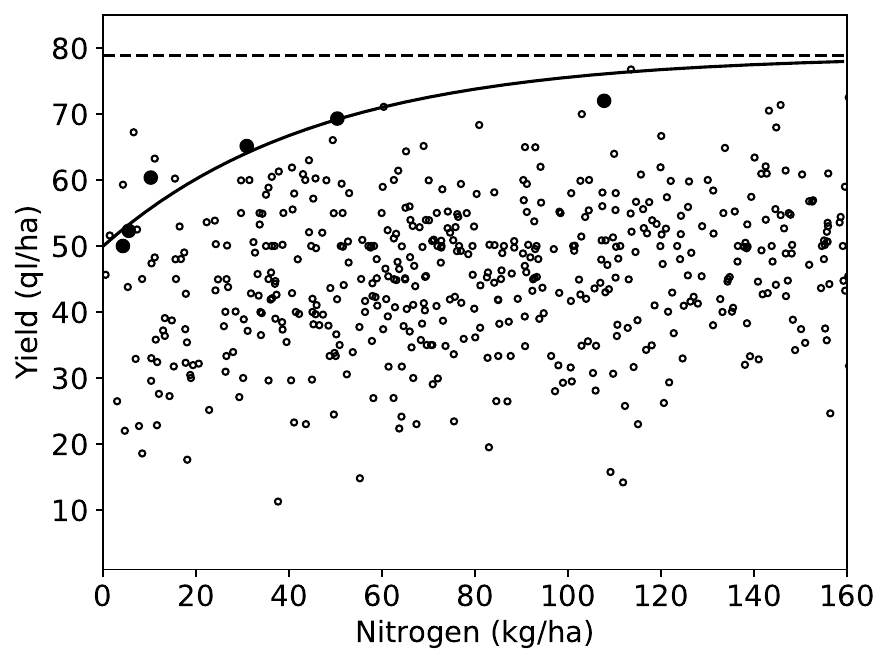}\hskip4mm
	\includegraphics[scale=0.4]{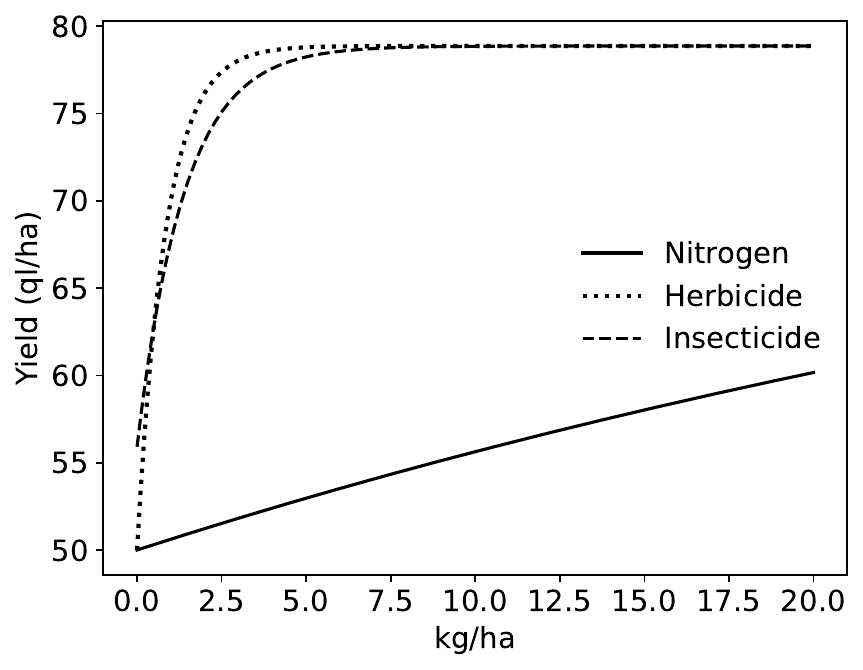}
	\caption{Visual representation of conditional yield curve estimation for nitrogen (left chart) for farms localized on the hills of the Macerata province. The same line is displayed, together with those estimated for herbicides and insecticides, in the right chart.}
	\label{fig:visual}
\end{figure}

The estimation process is based on equation (\ref{eq:condy}). We do not enter into the analytical details of the estimation. However, the overall strategy is as follows:
\begin{itemize}
\item Take the data of a set of farms
\item Identify the convex hull in the $x_i$-$y_i$ plane
\item Select points in the north-west portion of the convex hull (NWCH). If the NWCH has a limited number of points, remove the NWCH vertices from the original data and identify a new NWCH. 
\item Find the parameters by a least squares estimation using the NWCH points.
\end{itemize}
A visual representation of the estimation process and its results concerning the yield-nitrogen relationship is presented in Figure \ref{fig:visual} (left chart). 
It is a scatter plot of the yield and Nitrogen couples found in the farms of our database located in the Macerata province, operating at an altitude classified as hilly. In the figure, the small circles represent the data points, and the black bullets represent the NWCH set vertices. They serve to fit the nitrogen conditional yield curve (solid line), while the dashed line denotes the potential yield ($\bar{y}$). 
The chart on the right in Figure \ref{fig:visual} reports the fitted curves for Herbicides and insecticides in addition to that of Nitrogen already displayed in the left chart. 

\subsection{The model at work}\label{subsec:example}
The case reported here considers the three stress factors mentioned in section \ref{subsec:data}: plant nutrition, weeds, and insects.
They are relieved respectively by Nitrogen supply and the spread of herbicides and insecticides.

\begin{table}[!ht]
	\centering
\begin{tabular}{c r c c | c c c}
	\hspace*{2mm}$i$\hspace*{2mm}&{stress factor\hspace*{4mm}}&\hspace*{2mm}relieve element ($x$)\hspace*{2mm}&$p_x$&$s$&$\bar{s}$&$\lambda$\\
	\hline
	0&Lack of Nitrogen&supply of N (kg)&1.5&0.5&0.5&0.06\\
	1&weeds&herbicide (kg)&10&0.4&0.4&0.5\\
	2&insects&insecticide (kg)&10&0.3&0.3&0.7\\
	\hline
\end{tabular}
\caption{Stress factor, relieving elements and their prices. The last three columns report the estimates of model parameters.}
	\label{tab:stress}
\end{table}

The features of the example are reported in Table \ref{tab:stress}. In addition to the input prices reported in the table, we set the price of wheat at 300 per ton and the potential yield to 8.4 tons: $p_w=300$ and $\bar{y}=8.4$.

Implementing the process described in section \ref{subsec:model} we obtain the following solution: $\hat{x}_0^* = 50.13$ \hskip5mm $\hat{x}_1^* = 5.57$\hskip5mm $\hat{x}_2^* = 3.57$, and $\hat{y}^* = 7.8$.

In other words, the farmer uses 50.13 kg/ha of nitrogen fertilizer, 5.57 kg/ha of herbicide, and 3.57 kg/ha of insecticide. When any other factor discomfort the crop, this input mix gives the optimal target yield of 7.8 tons/ha.

This result represents the economically optimal input mix under standard agronomic conditions, assuming no additional biophysical constraints affect the crop. In the following section, we evaluate the environmental sustainability of this input combination using Life Cycle Assessment (LCA) techniques. This enables us to evaluate the trade-offs between economic and ecological performance, and to inform farmers of the environmental implications of their decisions.

\section{Impact assessment}\label{sec:lca}

\subsection{Life Cycle Assessment (LCA)}
In this study, the environmental impact of wheat production is assessed through the LCA analysis. The International Organization for Standardization (ISO) sets up a standard for this methodology (ISO 14040:2006). It recommends the analysis go through 4 phases: 1) goal and scope definition, 2) inventory analysis, 3) impact assessment, and 4) results interpretation. Phases 2) and 3) deserve particular attention and are briefly discussed hereafter.
The life cycle inventory analysis (LCI) involves recording all inputs and outputs associated with the production of the considered items. Inputs are distinguished into those coming from the environment and those coming from other production processes. Similarly, outputs are distinguished between those used in other production processes and those released into the environment (emissions). Inputs and outputs from and to other production processes are said to be technosphere items, while inputs and outputs from and to the environment are said to be biosphere items. Consider that the same input/output analysis can be applied to the technosphere inputs and outputs of the item being considered, and subsequently to the inputs of the inputs and the outputs of the outputs. Iterations move until a part or the whole life cycle of the considered item is covered. At each iteration, the set of biosphere items expands, and their quantities accumulate gradually. The impact assessment phase (LCIA) utilizes one or more assessment methods developed by the scientific community. An assessment method is a function that takes as inputs the quantities of biosphere items, i.e., the outputs of the inventory analysis, and transforms them into an indicator or a damaging substance directly linked to an environmental aspect. The transformation is made via coefficients studied and provided by the researchers who developed the method. In LCIA, these coefficients are more commonly referred to as characterization factors (CFs).
We will detail our choices for LCI and LCIA hereafter.
\subsection{LCI}
\textbf{LCI of fertilizers.}
The environmental effects of mineral fertilizers are analyzed in \cite{Isherwood1998}. Another interesting document concerning fertilizers is \cite{ifa2022}. The report just cited points out that Nitrogen is mainly responsible for emissions (see in particular chapter 1, paragraphs 6-11).
To integrate nitrogen fertilization into our LCA, we develop a process based on \cite{brentrup2004265} and, particularly, by referencing the median case in Appendix A of their paper.  

\noindent\textbf{LCI of Pesticides.}
LCI of pesticides is based on the active ingredients contained in the commercial products.
Unfortunately, the RICA database does not specify the name of the commercial product used by the farmer or the active ingredients.
To trace the information on active ingredients, we use the Fitogest$^{\textnormal{®}}$+ database, which provides detailed information on pesticides available in Italy. The database can be accessed at: \url{https://fitogest.imagelinenetwork.com/it/}.

Examining the most recurring ingredients in the available commercial products, we extract the information displayed in Table \ref{tab:pest_ap}. In particular, after analyzing the package leaflet of commercial products containing the identified active principle, we identify the range of active principle to be used per hectare (reported in the last column of table \ref{tab:pest_ap})

\begin{table}[h!]
\centering
\begin{tabular}{l c c c}
	type &\hspace*{2mm}toxicity level\hspace*{2mm}&\hspace*{2mm}active ingredient\hspace*{2mm} &  \hspace*{2mm}active ingredient per ha\hspace*{2mm}\\
\hline
herbicide & irritating & 2,4-D  & 360-720 g/ha \\
Insecticide & toxic & Pirimicarb & 130 g/ha \\
\hline
\end{tabular}
	\caption{Pesticides active principles}
	\label{tab:pest_ap}
\end{table}
\noindent\textbf{LCI of Tractor use.}
Because we aim to rely on open-source resources, the analysis is based on an adaptation of the LCI data provided by the Federal LCA Commons (\url{https://www.lcacommons.gov}). In particular, we refer to the University of Washington Design for Environment Laboratory/Field Crop Production database, which contains information on several processes related to the work of agricultural tractors for various crops in the US states. Among them, there is, for
example, a process named “work; ag. tractors for growing win wheat, 2014 fleet, all fuels; 100-175HP - US-AR” gathering inputs and outputs of an agriculture tractor producing 1 megajoule of work employed in winter wheat production in Arizona. In the lack of other freely available datasets, and as a first approximation, we will use the above mentioned dataset in this study.
Future work will aim to refine the input data, especially regarding active ingredient usage, mechanization profiles, and localized emission factors, to improve representativeness and robustness.

\subsection{LCIA}

\subsection{The ReCiPe methodology}

The recent literature on wheat LCA uses mainly the ReCiPe 2016 methodology.
\cite{JIANG2021143342} compare wheat production under different fertilizing strategies using ReCiPe 2016.
\cite{XIONG2024141423} perform a sustainability analysis of irrigated and rainfed wheat production systems under varying levels of nitrogen fertilizer using ReCiPe 2016.
\cite{DICRISTOFARO2024140696} Evaluate the impacts of different wheat farming systems through Life Cycle Assessment. They employ a method developed by \cite{resources8010056}, which, according to the author, is based on ReCiPe 2008.
\cite{YU2024171097} forecast environmental impacts of smallholder wheat production by coupling life cycle assessment and machine learning, using a selected number of impact assessments from different institutions including ReCiPe.

The ReCiPe 2016 methodology is described in \cite{Huijbregts2016,Huijbregts2017}.
Following the initial release in 2008, the ReCiPe methodology underwent an update in 2016. While the 2008 CFs concern the European scale, the 2016 version is representative at the global scale. Country and continental scale factors are provided for a number of impact categories.

ReCiPe provides impact factors both at Midpoint and at Endpoint. 
At the midpoint level, each method delivers a physical quantity, which in general is the most damaging substance for the considered category.
At the Endpoint level, each method is associated with one of three considered areas of protection: human health, ecosystem quality, and resource scarcity.
The damage to each of these three areas is measured as follows:
\begin{itemize}
	\item Damages to human health are measured by an indicator called ``Disability Adjusted Life Years'' (DALY) that gives the time (in years) that
		are lost, or that a person is disabled due to a disease or accident. 
	\item Damages to ecosystem quality are measured by the number of local species lost per year.
	\item Damages to resource scarcity are computed as the extra costs involved for future mineral and fossil resource extraction. It is expressed in Dollars.
\end{itemize}

In Endpoint analysis, each of the ReCiPe methods 
delivers a result expressed in one of these three units of measure.
This enables a nested aggregation process that identifies damages to specific subsystems. 

Both Midpoint and Endpoint impacts are provided, where possible, under three different perspectives, each with distinct features, the most significant of which is perhaps the time horizon. The three perspectives are called Individualist with a typical time horizon of 20 years, Hierarchist 100 years, and Egalitarian 500 years.

\subsection{ReCiPe adaptation}

\subsubsection{Regionalization.}

In ReCiPe 2016, the following methods have country-specific impact factors:
\begin{itemize}
	\item Fine dust formation
	\item Photochemical ozone formation - human health damage
	\item Photochemical ozone formation - ecosystem damage
	\item terrestrial acidification
	\item freshwater eutrophication
\end{itemize}

These methods include both global midpoint CFs and those specific to Italy.
Using the ratio between the values for Italy and those at the global level, we rescale all the CFs —i.e., endpoint and midpoint —under different perspectives to achieve a more accurate evaluation of the Italian case.

\subsubsection{Pesticides}

ReCiPe 2016 has CFs to perform the impact assessment of the active ingredients listed in Table \ref{tab:pest_ap}. Table \ref{tab:pests_in_bs3} reports, as an example, the CFs provided by the ReCiPe Ecotoxicity assessment method (Hierarchist perspective).

\begin{table}[!htb]
	\centering
\begin{tabular}{l l l l l l}
\hline
Type& tox level &Active Principle &  terrestrial &    freshwater   &  marine \\ 
\hline
Herbicide&irritating&2,4-D            &  0.042       &    0.359        &  0.02\\
Insecticide&toxic&Pirimicarb       &  0.378       &    0.455        &  0.038\\
\hline
\end{tabular}
	\caption{'ReCiPe 2016', '1.1 (20180117)', 'Midpoint', 'Ecotoxicity', 'Hierarchist' unit: kg 1,4-DCB equivalent (DCB=Dichlorobenzene)}
	\label{tab:pests_in_bs3}
\end{table}

\subsubsection{The set of LCIA methods}
After evaluating the possibility offered by our dataset, we decided to assess the impact of the following aspects:
\begin{itemize}
	\item	Terrestrial Acidification
               \item   Particulate Matter Formation
	       \item  		  Ozone Formation
                \item Freshwater Eutrophication
		\item Global Warming 100-year timescale
             	\item Toxicity
\end{itemize}

\begin{table}[!htb]
	\scriptsize
	\hskip-1cm
	\begin{tabular}{llllll}
	                   Method&                  Damage to &Geo CFs & Midpoint unit & Emitted to& Endpoint unit \\
\hline
Terrestrial Acidification&                 Ecosystems &  Italy &kg SO2-eq& soil&species.year\\
     Particulate Matter Formation&                     Humans &  Italy &kg PM2.5-eq&air&DALY\\
     		  Ozone Formation&                     Humans &  Italy &kg NOx-eq&air&DALY\\
                  \hskip7mm''&                 Ecosystems &  Italy &\hskip5mm''&air&species.year\\
        Freshwater Eutrophication&                 Ecosystems &  Italy &kg P-eq.       &freshwater&species.year\\
Global Warming 100 year timescale&       Humans and Ecoystems & Global &kg CO2-eq &air&DALY\\
			 Toxicity&      Humans - Carcinogenic & Global &kg 1,4-DCB eq. &urban air&DALY\\
			 \hskip3mm''&  Humans - Non-carcinog. & Global &\hskip5mm'' &urban air&DALY\\
                         \hskip3mm''&   Ecosystems - Terrestrial & Global &\hskip5mm'' &industrial soil&species.year\\
                         \hskip3mm''&    Ecosystems - Freshwater & Global &\hskip5mm'' &freshwater&species.year\\
\hline
		 \end{tabular}
	SO2=Sulfur dioxide; PM=Particle matter; NOx=Nitrogen Oxides; P=Phosporus; CO2=Carbon Dioxide; DCB=Dichlorobenzene
	\caption{Selected ReCiPe 2016 LCIA methods}
	\label{tab:selected_LCIA}
\end{table}
As reported in Table \ref{tab:selected_LCIA}, ozone formation comprises two methods: one to evaluate the impact on humans and the other to evaluate the environmental impact. Toxicity reaches a greater detail by distinguishing between carcinogenic and non-carcinogenic impacts on humans. Even the Ecosystem impact is further refined to evaluate terrestrial and Freshwater impacts. 
Table \ref{tab:selected_LCIA} also shows details on units of measure and regionalization. As explained above, we use the information in \cite{Huijbregts2016} to generate specific impact methods for Italy (the five methods in the top part of Table \ref{tab:selected_LCIA}). The other impact assessment methods are the ReCiPe 2016 originals.

\section{Results}\label{sec:results}
We are now equipped to evaluate the case reported in section \ref{subsec:example}. However, two adjustments are needed. The first one is for pesticides (herbicides and insecticides). It is because, as already mentioned, the RICA database does not include information on the active ingredients of pesticides. Therefore, we have to rely on the figures obtained from Fitogest. 

In particular, from Table \ref{tab:pest_ap}, we know that herbicide treatments often involve spreading 360-720 g/ha of the 2,4-D substance. Here, we will use the average value of 540 g/ha. Regarding insecticide, we have only the option of including 130g/ha of Pirimicarb.
The second adjustment concerns the use of tractor power, which we estimate at 900 MJ.

Summing up, we evaluate the impact of growing 1 hectare of wheat by an Italian farm using the following inputs:

\begin{itemize}
	\item 900MJ of tractor power
	\item 50kg of nitrogen for fertilization
	\item 0.54kg of 2,4-D for weeds control
	\item 0.13kg of Pirimicarb for insect control
\end{itemize}

Performing the LCA with these values, we get the results reported in table \ref{tab:LCA_results}.

\begin{table}[!htb]
	\scriptsize
	\hskip-1.5cm
	\begin{tabular}{lllllll}
	                   Method&                  Damage to &Geo CFs & Score & Unit & Score & Unit \\
\hline
Global Warming 100 year timescale&       Humans and Ecoystems & Global  & 943.1366&kg CO2-eq      & 8.752307e-04 &DALY\\
			 Toxicity&      Humans - Carcinogenic & Global  &   1.1561&kg 1,4-DCB-eq  & 3.838143e-06 &DALY\\
                         Toxicity&  Humans - Non-carcinogenic & Global  &   0.2679&kg 1,4-DCB-eq  & 6.108489e-08 &DALY\\
     Particulate Matter Formation&                     Humans &  Italy  &   2.6462&kg PM2.5-eq    & 6.250000e-11 &DALY\\
     		  Ozone Formation&                     Humans &  Italy  &   3.3371&kg NOx-eq      & 3.036748e-06 &DALY\\
		  \hline
	Terrestrial Acidification&                 Ecosystems &  Italy  &  15.9803&kg SO2-eq      & 3.387814e-06 &Species.year\\
                  Ozone Formation&                 Ecosystems &  Italy  &   7.6806&kg NOx-eq      & 9.908018e-07 &Species.year\\
        Freshwater Eutrophication&                 Ecosystems &  Italy  &   0.0166&kg P-eq        & 1.114726e-08 &Species.year\\
                         Toxicity&   Ecosystems - Terrestrial & Global  & 185.9196&kg 1,4-DCB-eq  & 2.119483e-09 &Species.year\\
                         Toxicity&    Ecosystems - Freshwater & Global  &   0.0606&kg 1,4-DCB-eq  & 4.211711e-11 &Species.year\\
			 \hline
	\end{tabular}
	\caption{Results of LCA performed on a farm with data available in RICA and evaluated as described above in the text.}
	\label{tab:LCA_results}
\end{table}

Using Table \ref{tab:LCA_results} we can assess that the most relevant damage to humans in terms of DALY caused by the farm we are investigating comes from the effect its activity has on climate change. 
The cultivation of wheat by this farm mostly impacts ecosystems via the terrestrial acidification, that has the largest impact of species loss per year. 

The figures for DALY and number of lost species are added to the agent information set. Therefore, agents are now aware of the impact of the traditional cultivation techniques on human health and the environment. The future development of the individual decision process is to include these values in the farmers' objective function.  We will therefore consider these values as ``influencing factors'' (\cite{HAYDEN202131}) that modify farmers' choices and increase the probability of leaving traditional production for more sustainable cultivation. 

\section{Toward an Agent-Based Model of the Italian Wheat Sector}

The final objective of this research is to build an agent-based model (ABM) of the Italian wheat production system.
We are developing the simulation code in Python using the Repast for Python framework (\url{https://repast.github.io/repast4py.site/index.html}). 
We include in our simulation a number of farms comparable to those operating in Italy (about 190000 according to the latest Italian agriculture census). 
To initialize the simulation, we use the statistical properties of farms found both in the RICA and in the latest Italian agriculture census databases (performed in 2020).

To account for market dynamics, the ABM will be interfaced with an existing simulation model (\cite{giulioni2019}) that delivers wheat prices in the most relevant international wheat markets. This allows us to make the price of wheat endogenous and to analyze the effects of relevant global shocks hindering the production of wheat in a specific area of the globe or its international trade. 

Each agent in the model represents a wheat-producing farm and is endowed with a behavioral module that includes i) a profit maximization function, based on the crop management model presented in Section 2; ii) an LCA module that computes environmental impacts using the ReCiPe 2016 Endpoint methodology, implemented through the open-source Brightway2 Python package (\url{https://docs.brightway.dev/en/legacy/index.html})

Relevant extensions will include behavioral elements and the effect of social interaction in adopting green practices.
Therefore, the results of the whole model will allow the evaluation of the introduction of sustainable policy, including those leveraging behavioral and social interaction elements (\cite{weersink}) aimed at fostering the appreciation of virtuous environmental practices by farmers. 

\vskip2mm
\noindent\textbf{Aknowledgements:} this research was conducted as part of the project ``ECOWHEATALY: Evaluation of policies for enhancing sustainable wheat production in Italy"  funded by the European Union-Next Generation EU under the call issued by the Minister of University and Research for the funding of research projects of relevant national interest (PRIN) grant n. PRIN 202288L9YN.

\bibliographystyle{chicago}
\bibliography{../references_ssc}

\begin{thebibliography}{}

\bibitem[\protect\citeauthoryear{Brentrup, Küsters, Lammel, Barraclough, and
  Kuhlmann}{Brentrup et~al.}{2004}]{brentrup2004265}
Brentrup, F., J.~Küsters, J.~Lammel, P.~Barraclough, and H.~Kuhlmann (2004).
\newblock {Environmental impact assessment of agricultural production systems
  using the life cycle assessment (LCA) methodology II. The application to N
  fertilizer use in winter wheat production systems}.
\newblock {\em European Journal of Agronomy\/}~{\em 20\/}(3), 265--279.

\bibitem[\protect\citeauthoryear{Carpentier, Gohin, Sckokai, and
  Thomas}{Carpentier et~al.}{2015}]{carpentier_atal_2015}
Carpentier, A., A.~Gohin, P.~Sckokai, and A.~Thomas (2015).
\newblock Economic modelling of agricultural production: past advances and new
  challenges.
\newblock {\em Review of agricultural and environmental studies\/}~{\em 96-1},
  131--165.

\bibitem[\protect\citeauthoryear{ClimaTalk}{ClimaTalk}{2024}]{yielgapclimatalk}
ClimaTalk (2024).
\newblock What is the yield gap?

\bibitem[\protect\citeauthoryear{Devkota, Bouasria, Devkota, and
  Nangia}{Devkota et~al.}{2024}]{DEVKOTA2024127195}
Devkota, K.~P., A.~Bouasria, M.~Devkota, and V.~Nangia (2024).
\newblock Predicting wheat yield gap and its determinants combining remote
  sensing, machine learning, and survey approaches in rainfed mediterranean
  regions of morocco.
\newblock {\em European Journal of Agronomy\/}~{\em 158}, 127195.

\bibitem[\protect\citeauthoryear{{di Cristofaro}, Marino, Lima, and
  Mastronardi}{{di Cristofaro} et~al.}{2024}]{DICRISTOFARO2024140696}
{di Cristofaro}, M., S.~Marino, G.~Lima, and L.~Mastronardi (2024).
\newblock Evaluating the impacts of different wheat farming systems through
  life cycle assessment.
\newblock {\em Journal of Cleaner Production\/}~{\em 436}, 140696.

\bibitem[\protect\citeauthoryear{{European Commission}}{{European
  Commission}}{2020}]{EU2020}
{European Commission} (2020).
\newblock Agri-environmental indicators.
\newblock
  \url{https://ec.europa.eu/eurostat/statistics-explained/index.php?title=Agri-environmental_indicators}.

\bibitem[\protect\citeauthoryear{Fischer}{Fischer}{2015}]{FISCHER20159}
Fischer, R. (2015).
\newblock Definitions and determination of crop yield, yield gaps, and of rates
  of change.
\newblock {\em Field Crops Research\/}~{\em 182}, 9--18.
\newblock SI:Yield potential.

\bibitem[\protect\citeauthoryear{{Food and Agriculture Organization of the
  United Nations}}{{Food and Agriculture Organization of the United
  Nations}}{2021}]{FAO2021}
{Food and Agriculture Organization of the United Nations} (2021).
\newblock The state of food and agriculture 2021.
\newblock \url{https://www.fao.org/publications/sofa/2021/en/}.

\bibitem[\protect\citeauthoryear{Giulioni, Di~Giuseppe, Toscano, Miglietta, and
  Pasqui}{Giulioni et~al.}{2019}]{giulioni2019}
Giulioni, G., E.~Di~Giuseppe, P.~Toscano, F.~Miglietta, and M.~Pasqui (2019).
\newblock A novel computational model of the wheat global market with an
  application to the 2010 russian federation case.
\newblock {\em Journal of Artificial Societies and Social Simulation\/}~{\em
  22\/}(3), 4.

\bibitem[\protect\citeauthoryear{Hayden, Mattimoe, and Jack}{Hayden
  et~al.}{2021}]{HAYDEN202131}
Hayden, M.~T., R.~Mattimoe, and L.~Jack (2021).
\newblock {Sensemaking and the influencing factors on farmer decision-making}.
\newblock {\em Journal of Rural Studies\/}~{\em 84}, 31--44.

\bibitem[\protect\citeauthoryear{Heady and Tweeten}{Heady and
  Tweeten}{1963}]{headyetal1963}
Heady, E.~O. and L.~G. Tweeten (1963).
\newblock {\em Resource Demand and Structure of the Agricultural Industry},
  Chapter Resource Substitutions in Agriculture.
\newblock Ames, IA: Iowa State University Press.

\bibitem[\protect\citeauthoryear{Huijbregts, Steinmann, Elshout, Stam, Verones,
  Vieira, Zijp, Hollander, and van Zelm}{Huijbregts
  et~al.}{2016}]{Huijbregts2016}
Huijbregts, M. A.~J., Z.~J.~N. Steinmann, P.~M.~F. Elshout, G.~Stam,
  F.~Verones, M.~Vieira, M.~Zijp, A.~Hollander, and R.~van Zelm (2016).
\newblock {ReCiPe 2016 A harmonized life cycle impact assessment method at
  midpoint and endpoint level Report I: Characterization}.
\newblock Technical Report {RIVM 2016-0104}, Dutch National Institute for
  Public Health and the Environment, Bilthoven, The Netherlands.
\newblock \url{https://www.rivm.nl/bibliotheek/rapporten/2016-0104.pdf}.

\bibitem[\protect\citeauthoryear{Huijbregts, Steinmann, Elshout, Stam, Verones,
  Vieira, Zijp, Hollander, and van Zelm}{Huijbregts
  et~al.}{2017}]{Huijbregts2017}
Huijbregts, M. A.~J., Z.~J.~N. Steinmann, P.~M.~F. Elshout, G.~Stam,
  F.~Verones, M.~Vieira, M.~Zijp, A.~Hollander, and R.~van Zelm (2017, Feb).
\newblock Recipe2016: a harmonised life cycle impact assessment method at
  midpoint and endpoint level.
\newblock {\em The International Journal of Life Cycle Assessment\/}~{\em
  22\/}(2), 138--147.

\bibitem[\protect\citeauthoryear{IFA and Systemiq}{IFA and
  Systemiq}{2022}]{ifa2022}
IFA and Systemiq (2022).
\newblock Reducing emissions from fertilizers use.
\newblock Technical report, Systemiq for International Fertilizer Industry
  Association (IFA).
\newblock \url{https://www.systemiq.earth/reducing-emissions-fertilizer/}.

\bibitem[\protect\citeauthoryear{{Intergovernmental Panel on Climate
  Change}}{{Intergovernmental Panel on Climate Change}}{2021}]{IPCC2021}
{Intergovernmental Panel on Climate Change} (2021).
\newblock {\em Climate Change 2021: The Physical Science Basis}.
\newblock Cambridge University Press.
\newblock \url{https://www.ipcc.ch/report/ar6/wg1/}.

\bibitem[\protect\citeauthoryear{Isherwood}{Isherwood}{1998}]{Isherwood1998}
Isherwood, K.~F. (1998).
\newblock Mineral fertilizer use and the environment.
\newblock Technical report, International Fertilizer Industry Association (IFA)
  and United Nations Environment Program (UNEP).
\newblock \url{https://digitallibrary.un.org/record/468871}.

\bibitem[\protect\citeauthoryear{Jiang, Zheng, and Xing}{Jiang
  et~al.}{2021}]{JIANG2021143342}
Jiang, Z., H.~Zheng, and B.~Xing (2021).
\newblock Environmental life cycle assessment of wheat production using
  chemical fertilizer, manure compost, and biochar-amended manure compost
  strategies.
\newblock {\em Science of The Total Environment\/}~{\em 760}, 143342.

\bibitem[\protect\citeauthoryear{Kay, Edwards, and Duffy}{Kay
  et~al.}{2020}]{kayetal2020}
Kay, R.~D., W.~M. Edwards, and P.~A. Duffy (2020).
\newblock {\em Farm management\/} (Ninth edition ed.).
\newblock McGraw-Hill Education.

\bibitem[\protect\citeauthoryear{Kunz}{Kunz}{2022}]{farmmanweb}
Kunz, K. (2022).
\newblock A complete guide to farm management.

\bibitem[\protect\citeauthoryear{Pretty, Benton, Bharucha, Dicks, Flora,
  Godfray, and Wratten}{Pretty et~al.}{2018}]{Pretty2018}
Pretty, J., T.~G. Benton, Z.~P. Bharucha, L.~V. Dicks, C.~B. Flora, H.~C.~J.
  Godfray, and S.~D. Wratten (2018).
\newblock Global assessment of agricultural system redesign for sustainable
  intensification.
\newblock {\em Nature Sustainability\/}~{\em 1\/}(8), 441--446.

\bibitem[\protect\citeauthoryear{Recchia, Cappelli, Cini, Garbati~Pegna, and
  Boncinelli}{Recchia et~al.}{2019}]{resources8010056}
Recchia, L., A.~Cappelli, E.~Cini, F.~Garbati~Pegna, and P.~Boncinelli (2019).
\newblock Environmental sustainability of pasta production chains: An
  integrated approach for comparing local and global chains.
\newblock {\em Resources\/}~{\em 8\/}(1).

\bibitem[\protect\citeauthoryear{Snyder, Bruulsema, Jensen, and Fixen}{Snyder
  et~al.}{2009}]{Snyder2009}
Snyder, C.~S., T.~W. Bruulsema, T.~L. Jensen, and P.~E. Fixen (2009).
\newblock Review of greenhouse gas emissions from crop production systems and
  fertilizer management effects.
\newblock {\em Agriculture, Ecosystems \& Environment\/}~{\em 133\/}(3--4),
  247--266.

\bibitem[\protect\citeauthoryear{Tilman, Cassman, Matson, Naylor, and
  Polasky}{Tilman et~al.}{2002}]{Tilman2002}
Tilman, D., K.~G. Cassman, P.~A. Matson, R.~Naylor, and S.~Polasky (2002).
\newblock Agricultural sustainability and intensive production practices.
\newblock {\em Nature\/}~{\em 418}, 671--677.

\bibitem[\protect\citeauthoryear{{van Ittersum}, Cassman, Grassini, Wolf,
  Tittonell, and Hochman}{{van Ittersum} et~al.}{2013}]{VANITTERSUM20134}
{van Ittersum}, M.~K., K.~G. Cassman, P.~Grassini, J.~Wolf, P.~Tittonell, and
  Z.~Hochman (2013).
\newblock Yield gap analysis with local to global relevance—a review.
\newblock {\em Field Crops Research\/}~{\em 143}, 4--17.
\newblock Crop Yield Gap Analysis – Rationale, Methods and Applications.

\bibitem[\protect\citeauthoryear{Weersink and Fulton}{Weersink and
  Fulton}{}]{weersink}
Weersink, A. and M.~Fulton.
\newblock Limits to profit maximization as a guide to behavior change.
\newblock {\em Applied Economic Perspectives and Policy\/}~{\em 42\/}(1),
  67--79.

\bibitem[\protect\citeauthoryear{Xiong, Shah, Zhao, Li, Zha, Ye, and Wu}{Xiong
  et~al.}{2024}]{XIONG2024141423}
Xiong, L., F.~Shah, Y.~Zhao, Z.~Li, X.~Zha, M.~Ye, and W.~Wu (2024).
\newblock Sustainability analysis of irrigated and rainfed wheat production
  systems under varying levels of nitrogen fertilizer through coupling of
  emergy accounting and life cycle assessment.
\newblock {\em Journal of Cleaner Production\/}~{\em 447}, 141423.

\bibitem[\protect\citeauthoryear{Yu, Xu, Cai, Li, Wang, Zhang, and Lin}{Yu
  et~al.}{2024}]{YU2024171097}
Yu, C., G.~Xu, M.~Cai, Y.~Li, L.~Wang, Y.~Zhang, and H.~Lin (2024).
\newblock Predicting environmental impacts of smallholder wheat production by
  coupling life cycle assessment and machine learning.
\newblock {\em Science of The Total Environment\/}~{\em 921}, 171097.

\end{thebibliography}
\end{document}